\documentclass[twocolumn,showpacs,aps,prl,superscriptaddress,letterpaper]{revtex4}

\usepackage{graphicx}
\usepackage{dcolumn}
\usepackage{amsmath}
\usepackage{epsfig}
\usepackage{multirow}

\long\def\inst#1{\par\nobreak\kern 4pt\nobreak
    {\itshape #1}\par\vskip 10pt plus 3pt minus 3pt}
\RequirePackage{xspace}
\usepackage{relsize}

\def\babar{\mbox{\slshape B\kern-0.1em{\smaller A}\kern-0.1em
    B\kern-0.1em{\smaller A\kern-0.2em R}}}
\def\Abar    {\kern 0.18em\overline{\kern -0.18em A}{}\xspace}
\def\Kbar    {\kern 0.18em\overline{\kern -0.18em K}{}\xspace}
\def\Bbar    {\kern 0.18em\overline{\kern -0.18em B}{}\xspace}

\def\BB      {\ensuremath{B\Bbar}\xspace} 
\def\Bz      {\ensuremath{B^0}\xspace}
\def\Bzb     {\ensuremath{\Bbar^0}\xspace}
\def\BzBzb   {\ensuremath{\Bz {\kern -0.16em \Bzb}}\xspace}
\def\Bu      {\ensuremath{B^+}\xspace}
\def\Bub     {\ensuremath{B^-}\xspace}

\def\BpBm    {\ensuremath{\Bu {\kern -0.16em \Bub}}\xspace}

\newcommand{\optbar}[1]{\shortstack{{\tiny (\rule[.4ex]{1em}{.1mm})}
  \\ [-.7ex] $#1$}}
\def\BorBbar    {\kern 0.18em\optbar{\kern -0.18em B}{}\xspace}
\def\DorDbar    {\kern 0.18em\optbar{\kern -0.18em D}{}\xspace}
\def\KorKbar    {\kern 0.18em\optbar{\kern -0.18em K}{}\xspace}
\def\CP                {\ensuremath{C\!P}\xspace}
\def\pep2{PEP-II}
\mathchardef\Upsilon="7107
\def\Y#1S{\ensuremath{\Upsilon{(#1S)}}\xspace}

\def\FourS {\Y4S}

\newcommand{\BABARPubYear}     {06}
\newcommand{\BABARPubNumber}  {064}

\newcommand{\SLACPubNumber} {12158}

\begin{document}

\begin{flushleft}
\babar-PUB-\BABARPubYear/\BABARPubNumber\\
SLAC-PUB-\SLACPubNumber
\\
hep-ex/0610073
\\[5mm]
\end{flushleft}

\title{
\large \bfseries \boldmath
Vector-Tensor and Vector-Vector Decay Amplitude Analysis of $B^0\to\phi K^{*0}$ 
}

%
\author{B.~Aubert}
\author{M.~Bona}
\author{D.~Boutigny}
\author{F.~Couderc}
\author{Y.~Karyotakis}
\author{J.~P.~Lees}
\author{V.~Poireau}
\author{V.~Tisserand}
\author{A.~Zghiche}
\affiliation{Laboratoire de Physique des Particules, IN2P3/CNRS et Universit\'e de Savoie, F-74941 Annecy-Le-Vieux, France }
\author{E.~Grauges}
\affiliation{Universitat de Barcelona, Facultat de Fisica, Departament ECM, E-08028 Barcelona, Spain }
% ...
\collaboration{The \babar\ Collaboration}
\noaffiliation

\date{October 24, 2006}

\begin{abstract}
We perform an amplitude analysis of the decays $B^0\to\phi K^{*}_2(1430)^0$, 
$\phi K^{*}(892)^0$, and $\phi(K\pi)^0_{S-{\rm wave}}$ with a sample of about 
384 million $\BB$ pairs recorded with the $\babar$ detector. The fractions of 
longitudinal polarization ${f_L}$ of the vector-tensor and vector-vector decay 
modes are measured to be $0.853^{+0.061}_{-0.069}\pm 0.036$ and 
$0.506\pm{0.040}\pm 0.015$, respectively. Overall, twelve parameters are measured 
for the vector-vector decay and seven parameters for the vector-tensor decay,
including the branching fractions and parameters sensitive to $C\!P$-violation.
\end{abstract}


\maketitle

\begin{figure}[b]
\centerline{\setlength{\epsfxsize}{1.0\linewidth}\leavevmode\epsfbox{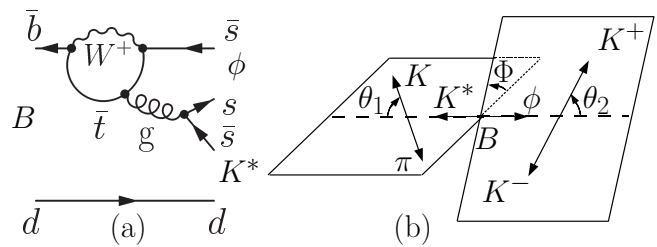}}
\caption{\label{fig:decay}
(a) Feynman diagram describing the $B^0\to\phi K^{*0}$ decay;
(b) definition of decay angles given in the rest frames of the decaying parents.
}
\end{figure}

The interest in the polarization and $C\!P$-asymmetry measurements in
$B\to\phi K^*$ decays is motivated by their potential sensitivity to 
physics beyond the standard model in the $b\to s$ transition, shown in 
Fig.~\ref{fig:decay}~(a)~\cite{bvv1}.
The polarization measurements of $B$ meson decays reveal both strong
and weak interaction dynamics and are discussed in a recent 
review~\cite{bvvreview2006,pdg2006}. The large 
fraction of transverse polarization in the $B\to\phi K^*(892)$ decay 
measured by \babar~\cite{babar:vv} and by Belle~\cite{belle:phikst} 
indicates a significant departure from the naive expectation of 
predominant longitudinal polarization. This suggests other contributions 
to the decay amplitude, previously neglected, either within or beyond 
the standard model~\cite{newtheory}. 

We now extend our investigation 
of the polarization puzzle with an amplitude analysis of the vector-tensor 
$B^0\to\phi K^{*}_2(1430)^0$ decay. We also measure vector-vector 
$B^0\to\phi K^{*}(892)^0$ and vector-scalar $B^0\to\phi(K\pi)_0^{*0}$ 
decay amplitudes, where $(K\pi)_0^{*0}$ is the $J^P=0^+$ $K\pi$ component.
We use the dependence on the $K\pi$ invariant mass of the
interference between the $J^P=0^+$ and $1^-$ or $2^+$
components~\cite{Aston:1987ir, jpsikpi}
to resolve the discrete ambiguity in the determination of the strong 
and weak phases otherwise present in the $B^0\to\phi K^{*}(892)^0$ 
analysis~\cite{bvvreview2006,babar:vv, belle:phikst} 
and to provide new measurements of the strong and 
weak phases relative to the vector-scalar decay amplitude.

The angular distribution of the $B\to\phi K^*$ decay can be expressed 
as a function of ${\cal H}_i=\cos\theta_i$ and $\Phi$ shown in 
Fig.~\ref{fig:decay}~(b). Here $\theta_i$ is the angle between the direction
of the $K$ meson from the $K^*\to K\pi$ ($\theta_1$) or $\phi\to K\Kbar$ 
($\theta_2$) and the direction opposite the $B$ in the $K^*$ or $\phi$
rest frame, and $\Phi$ is the angle between the decay planes of the two 
systems. The differential decay width has seven complex amplitudes 
$A_{J\lambda}$ corresponding to the spin of the $K\pi$ system $J$ and
the helicity $\lambda=0$ or $\pm 1$:
\begin{eqnarray}
\label{eq:helicityfull}
{d^3\Gamma \over d{\cal H}_1 d{\cal H}_2d\Phi} \propto
\left|~\sum_{} 
A_{J\lambda} Y_{J}^{\lambda}({\cal H}_1,\Phi) Y_{1}^{-\!\lambda}(-{\cal H}_2,0)~\right|^2,
\end{eqnarray}
where $Y_{J}^{\lambda}$ are the spherical harmonics with $J=2$ for 
$K_2^{*}(1430)$, $J=1$ for $K^{*}(892)$, and $J=0$ for $(K\pi)_0^{*}$.
We can reparameterize the amplitudes with the index $J$ suppressed as  
$A_0$ and $A_{\pm1}=(A_{\parallel}\pm A_{\perp})/\sqrt{2}$.

We analyze $\BorBbar^0\to\phi\KorKbar^{*0}\to(K^+K^-)({K^\pm\pi^\mp})$
candidates using data collected with the \babar\ detector~\cite{babar} 
at the PEP-II $e^+e^-$ collider. A sample of $383.6\pm 4.2$ 
million $\FourS\to\BB$ events was recorded at the center-of-mass  
energy $\sqrt{s} = 10.58$ GeV. Charged-particle momenta are measured 
in a tracking system consisting of a silicon vertex tracker with five 
double-sided layers and a 40-layer drift chamber, both within the 1.5-T 
magnetic field of a solenoid. Charged-particle identification is provided 
by measurements of the energy loss in the tracking devices and by 
a ring-imaging Cherenkov detector.


We use two kinematic variables:
$\Delta{E}=(E_iE_B-\mathbf{p}_i\cdot\mathbf{p}_B-s/2)/\sqrt{s}$
and $m_{\rm{ES}} = [{ (s/2 + \mathbf{p}_i \cdot 
\mathbf{p}_B)^2 / E_i^2 - \mathbf{p}_B^{\,2} }]^{1/2}$,
where $(E_i,\mathbf{p}_i)$ is the $e^+e^-$ beam four-momentum, 
and $(E_B,\mathbf{p}_B)$ is the four-momentum of the $B$ candidate.
We require $|\Delta{E}|<0.1$ GeV and $m_{\rm{ES}}>5.25$ GeV.
The requirements on the invariant masses are 
$0.99 < m_{K\!\Kbar} < 1.05$ GeV and 
$0.75 < m_{K\!\pi} < 1.05$ GeV (lower $m_{K\!\pi}$ range)
or $1.13 < m_{K\!\pi} < 1.53$ GeV (higher $m_{K\!\pi}$ range).

To reject the dominant $e^+e^-\to$ quark-antiquark background, 
we use variables calculated in the center-of-mass frame.
We require $|\cos\theta_T| < 0.8$, where $\theta_T$ 
is the angle between the $B$-candidate thrust axis
and that of the rest of the event.
We construct a Fisher discriminant, ${\cal F}$,
that combines the polar angles of the $B$-momentum vector 
and the $B$-candidate thrust axis with respect to the beam axis,
and the two Legendre moments $L_0$ and $L_2$ of the energy 
flow around the $B$-candidate thrust axis~\cite{bigPRD}.

We remove signal candidates that have decay products 
with invariant mass within 12 MeV of the nominal mass 
values for $D_s^\pm$ or $D^\pm\to{\phi\pi^\pm}$.
In about $5\%$ of events, more than one candidate is 
reconstructed, and we select the one whose four-track 
vertex has the lowest $\chi^2$. We define the flavor 
sign $Q$ to be the charge of the pion.


We use an unbinned, extended maximum-likelihood fit~\cite{babar:vv} 
to extract the event yields $n_j$, flavor asymmetries ${\cal A}_j$, 
and the probability density function (PDF) parameters, denoted by
{\boldmath$\zeta$} for the polarization parameters and {\boldmath$\xi$}
for the remaining parameters.
The data model has five event categories $j$: 
$B\to\phi(K\pi)_{J=0,1,2}$, $B\to f_0(980)K^*$,
and combinatorial background.
The combinatorial background PDF is found to account 
well for both the dominant quark-antiquark background 
and the random tracks from the $B$ decays.
The likelihood ${\cal L}_i$ for each candidate $i$ is defined as
${\cal L}_i = \sum_{j,k}n_{j}^k\, 
{\cal P}_{j}^k$({\boldmath ${\rm x}_i$};~{\boldmath$\zeta$};~{\boldmath$\xi$}),
where each of the ${\cal P}_{j}^k$ is the PDF for variables
{\boldmath ${\rm x}_i$}~$=\{{\cal H}_1$, ${\cal H}_2$, $\Phi$, 
$m_{K\!\pi}$, $m_{K\!\Kbar}$, $\Delta E$, $m_{\rm{ES}}$, ${\cal F}$, $Q$\}.
The flavor index $k$ corresponds to the value of $Q$, that is
${\cal P}_{j}^k\equiv{\cal P}_{j}\times\delta_{kQ}$. 

We define
$n_{j}=n^+_{j}+n^-_{j}$ and ${\cal A}_j=(n^+_{j}-n^-_{j})/(n^+_{j}+n^-_{j})$.
The polarization parameters, with the index $J$ suppressed, 
are defined as $f_L={|A_0|^2/\Sigma|A_\lambda|^2}$,
$f_{\perp}={|A_{\perp}|^2/\Sigma|A_\lambda|^2}$,
$\phi_{\parallel} = {\rm arg}(A_{\parallel}/A_0)$, and
$\phi_{\perp} = {\rm arg}(A_{\perp}/A_0)$.
We allow for $C\!P$-violating differences between the $\Bbar^0$ 
($Q=+1$) and ${B}^0$ ($Q=-1$) decay amplitudes ($\Abar$ and $A$)
and incorporate them via the replacements 
$f_L\to f_L\times(1+{\cal A}^0_{C\!P}\times Q)$, 
$f_\perp\to f_\perp\times(1+{\cal A}^\perp_{C\!P}\times Q)$, 
$\phi_\parallel\to(\phi_\parallel+\Delta\phi_\parallel\times Q)$, and
$\phi_\perp\to(\phi_\perp+{\pi/2}+(\Delta\phi_\perp+{\pi/2})\times Q)$~\cite{bvvreview2006}.

The PDF 
${\cal P}_{j}$({\boldmath ${\rm x}_i$};~{\boldmath$\zeta$};~{\boldmath$\xi$}) 
for a given candidate $i$ is a joint PDF for the helicity angles, 
resonance masses, and $Q$, and the product of 
the PDFs for each of the remaining variables.
The helicity part of the exclusive $B$ decay PDF is the 
ideal angular distribution from Eq.~(\ref{eq:helicityfull}),
where the amplitudes $A_{J\lambda}$ are expressed in terms of
the polarization parameters {\boldmath$\zeta$}, multiplied 
by an empirically-determined acceptance function
${\cal{G}}({\cal H}_1,{\cal H}_2,\Phi)
\equiv{\cal{G}}_1({\cal H}_1)\times{\cal{G}}_2({\cal H}_2)$.
A relativistic $J$-spin Breit-Wigner amplitude
parameterization is used for the resonance 
mass~\cite{pdg2006,f0mass}, 
except for the $(K\pi)^{*0}_0$ $m_{K\!\pi}$ amplitude 
parameterized with the LASS function~\cite{Aston:1987ir}.
The latter includes the $K_0^{*}(1430)^0$ resonance
together with a nonresonant component.

\begin{figure}[t]
\centerline{
\setlength{\epsfxsize}{0.5\linewidth}\leavevmode\epsfbox{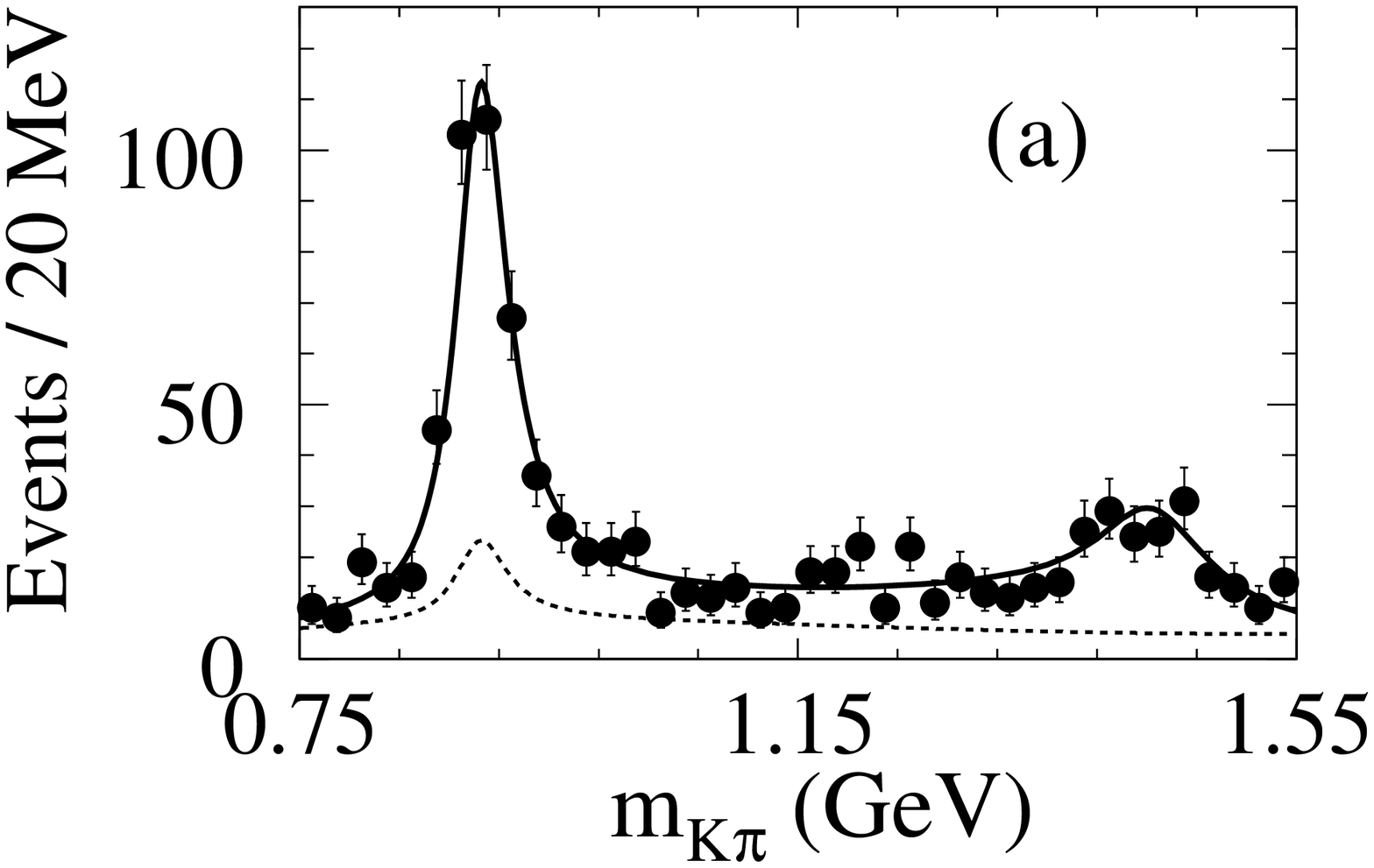}
\setlength{\epsfxsize}{0.5\linewidth}\leavevmode\epsfbox{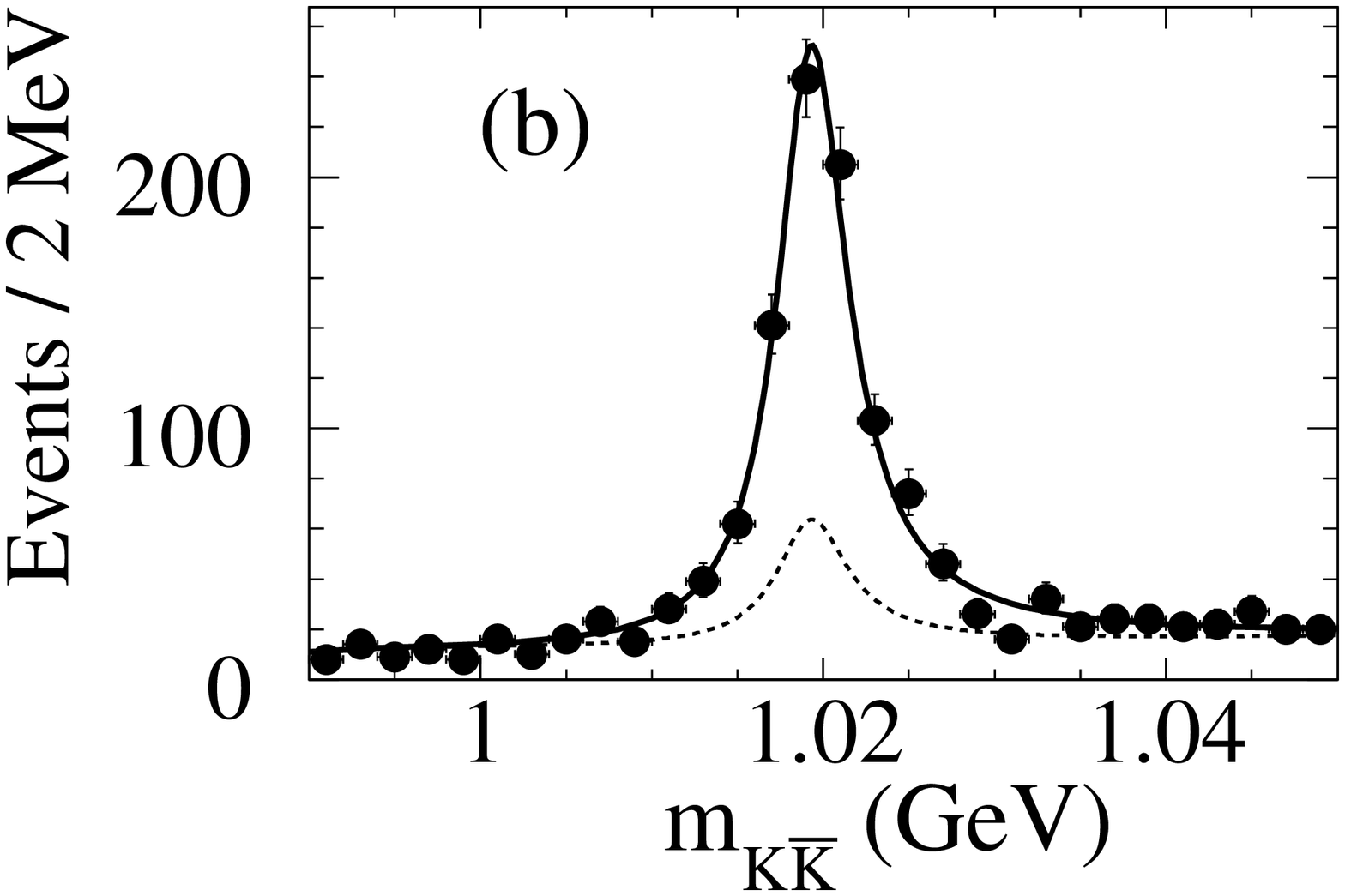}
}
\centerline{
\setlength{\epsfxsize}{0.5\linewidth}\leavevmode\epsfbox{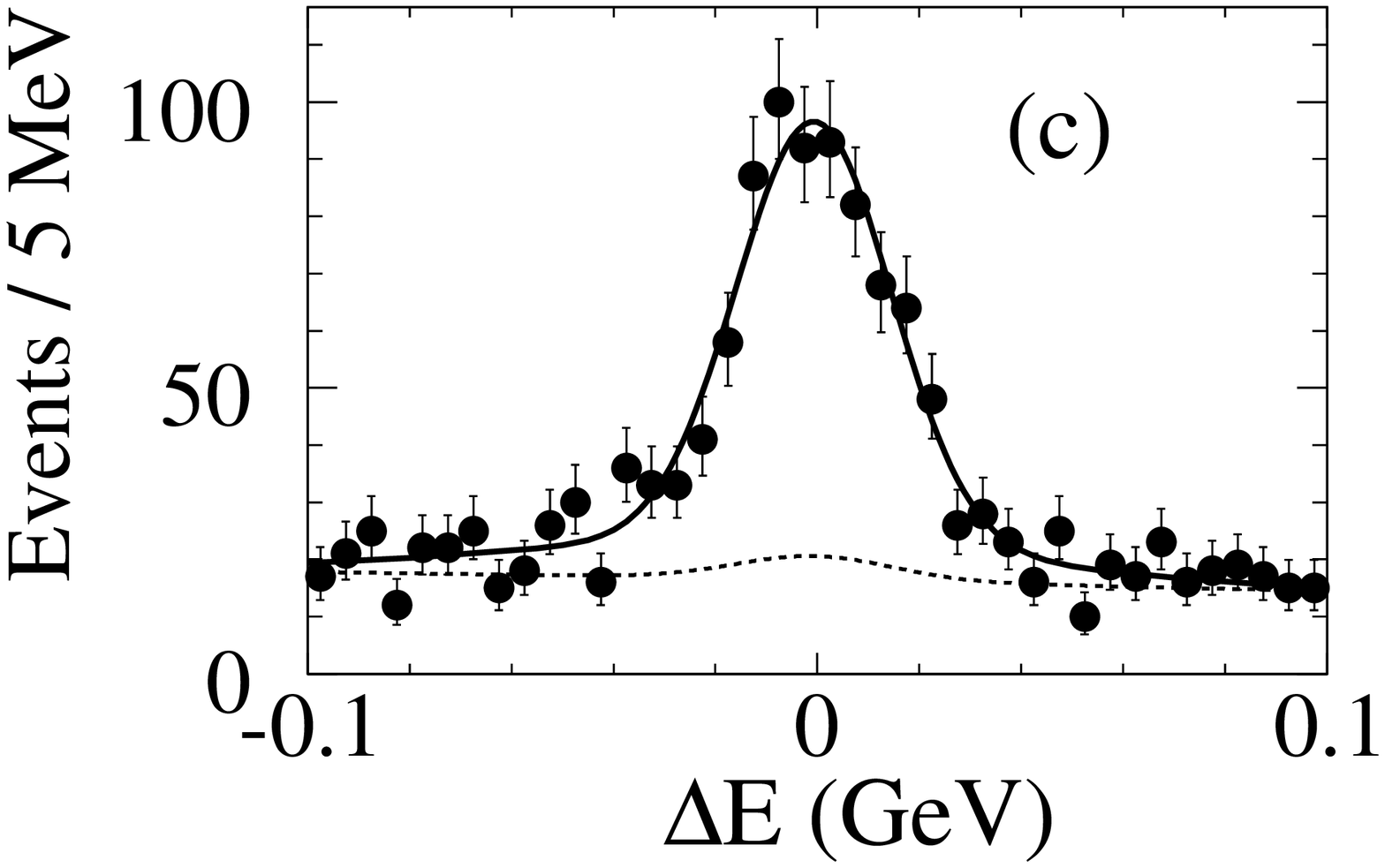}
\setlength{\epsfxsize}{0.5\linewidth}\leavevmode\epsfbox{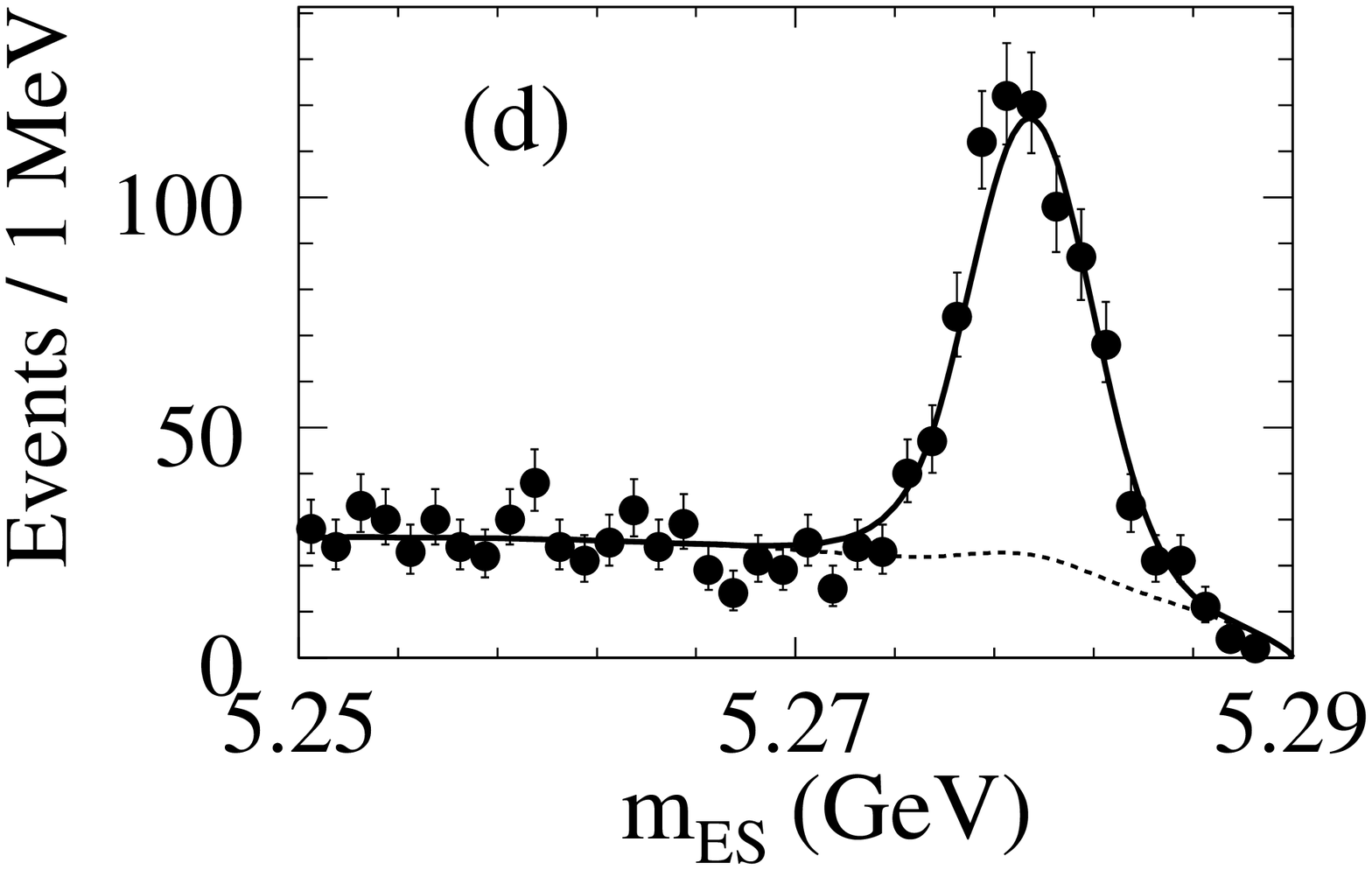}
}
\vspace{-0.3cm}
\caption{\label{fig:projection1} 
Projections onto the variables  $m_{K\!\pi}$ (a), $m_{K\!\Kbar}$ (b),  $\Delta E$ (c),
and $m_{\rm ES}$ (d) for the signal $B^0\to\phi(K\pi)$ candidates.
Data distributions are shown with a requirement on the signal-to-background
probability ratio calculated with the plotted variable excluded.
The solid (dashed) lines show the signal-plus-background
(background) PDF projections.
}
\end{figure}
\begin{figure}[b]
\centerline{
\setlength{\epsfxsize}{0.5\linewidth}\leavevmode\epsfbox{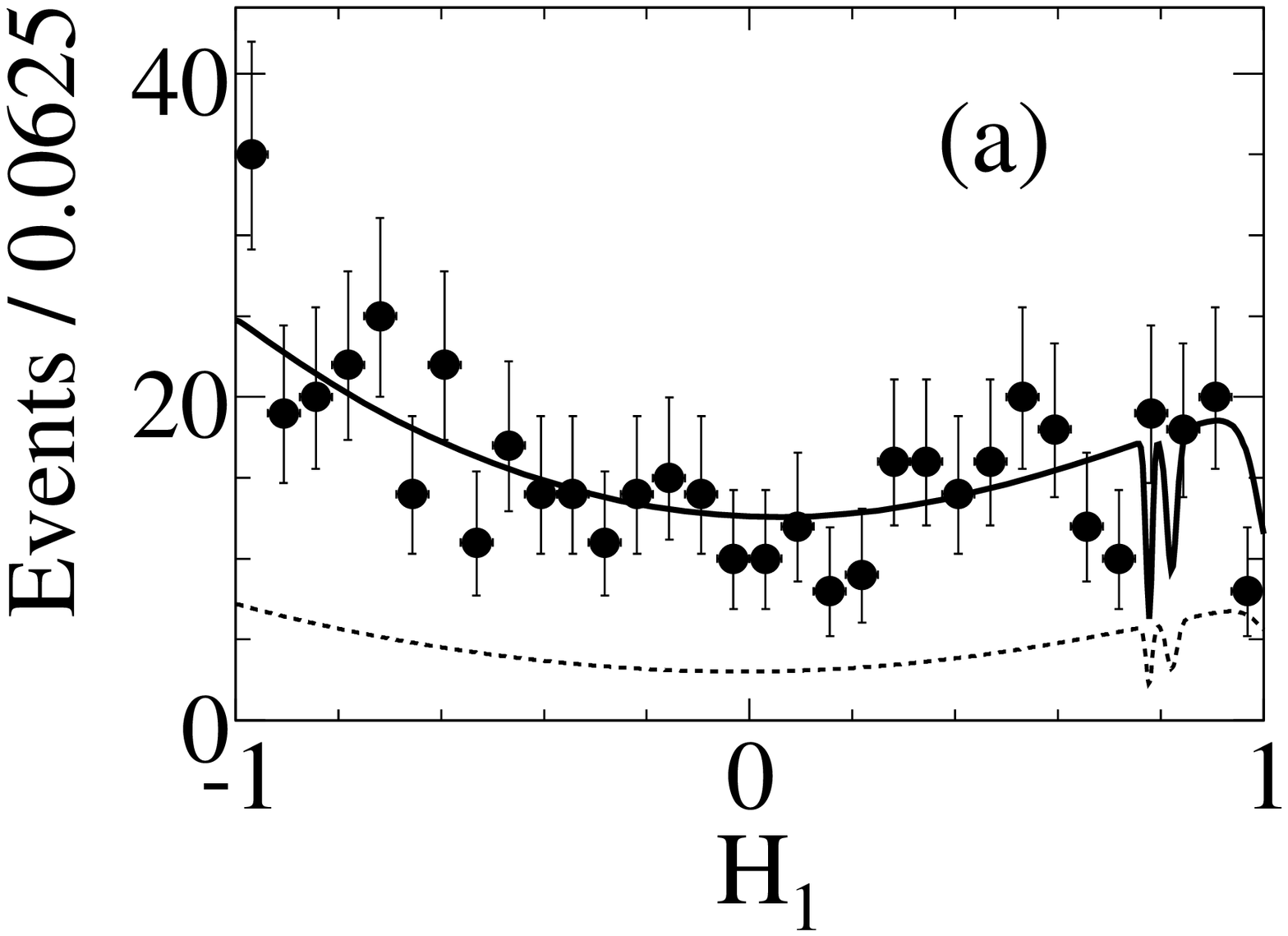}
\setlength{\epsfxsize}{0.5\linewidth}\leavevmode\epsfbox{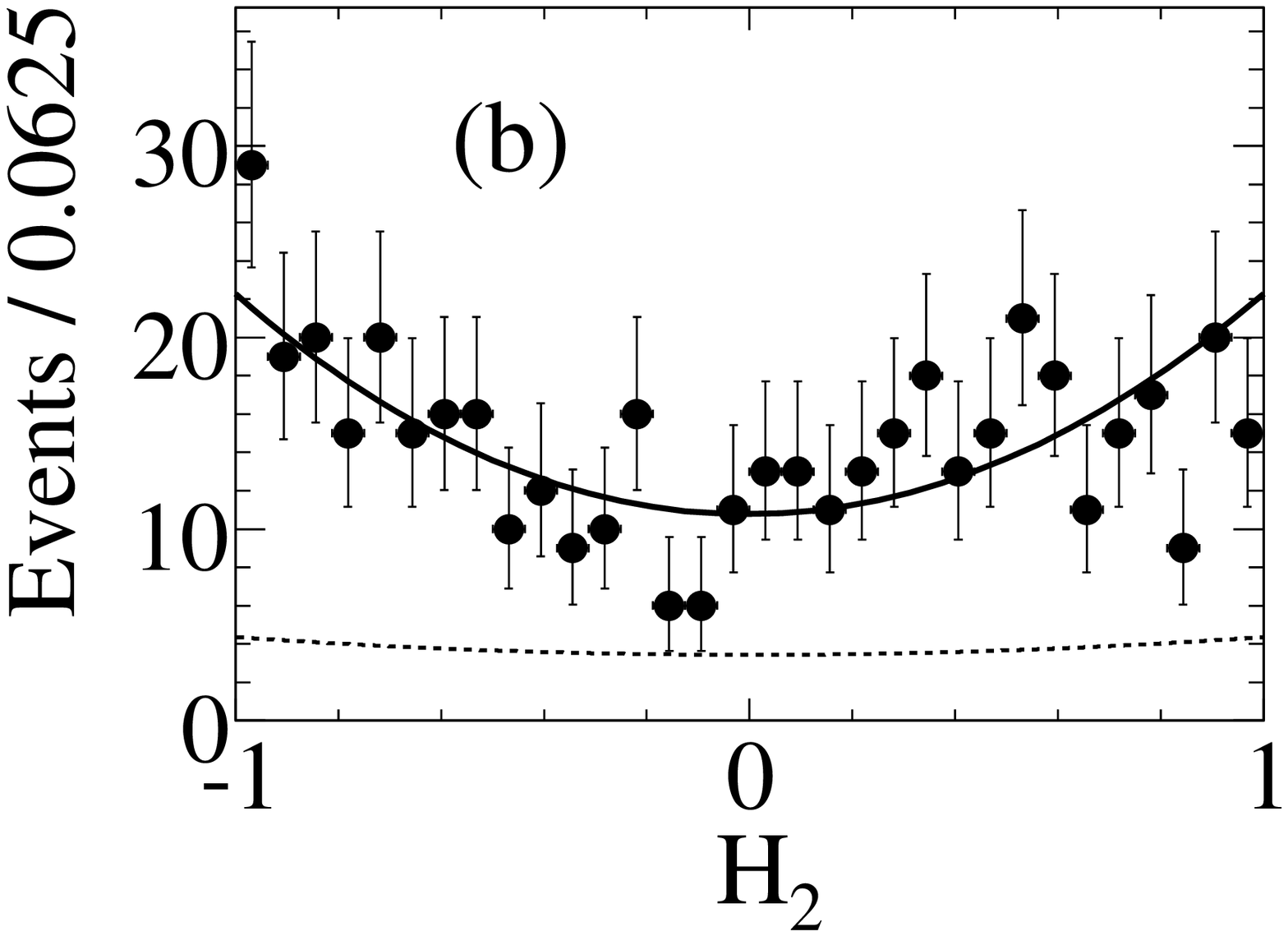}
}
\centerline{
\setlength{\epsfxsize}{0.5\linewidth}\leavevmode\epsfbox{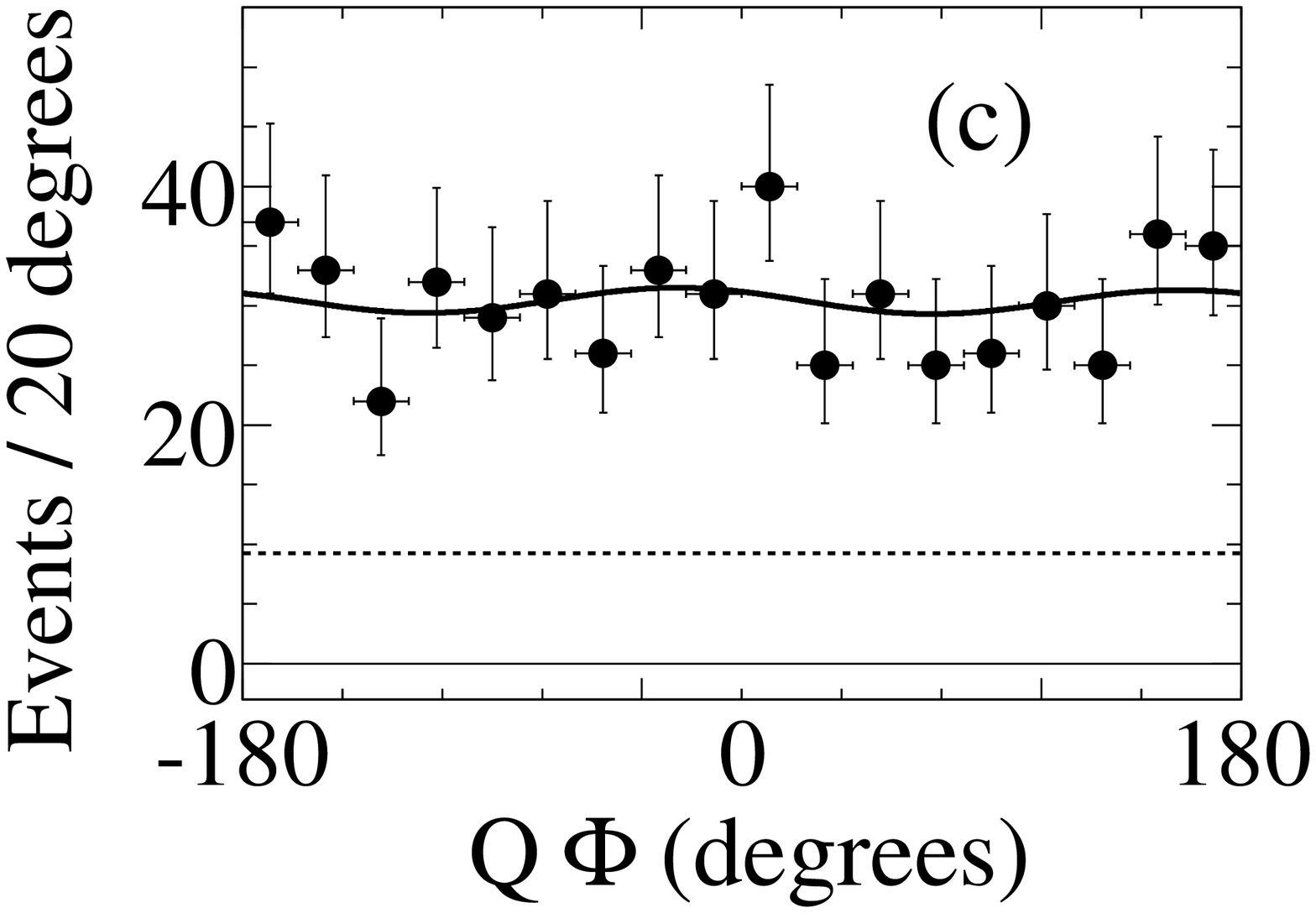}
\setlength{\epsfxsize}{0.5\linewidth}\leavevmode\epsfbox{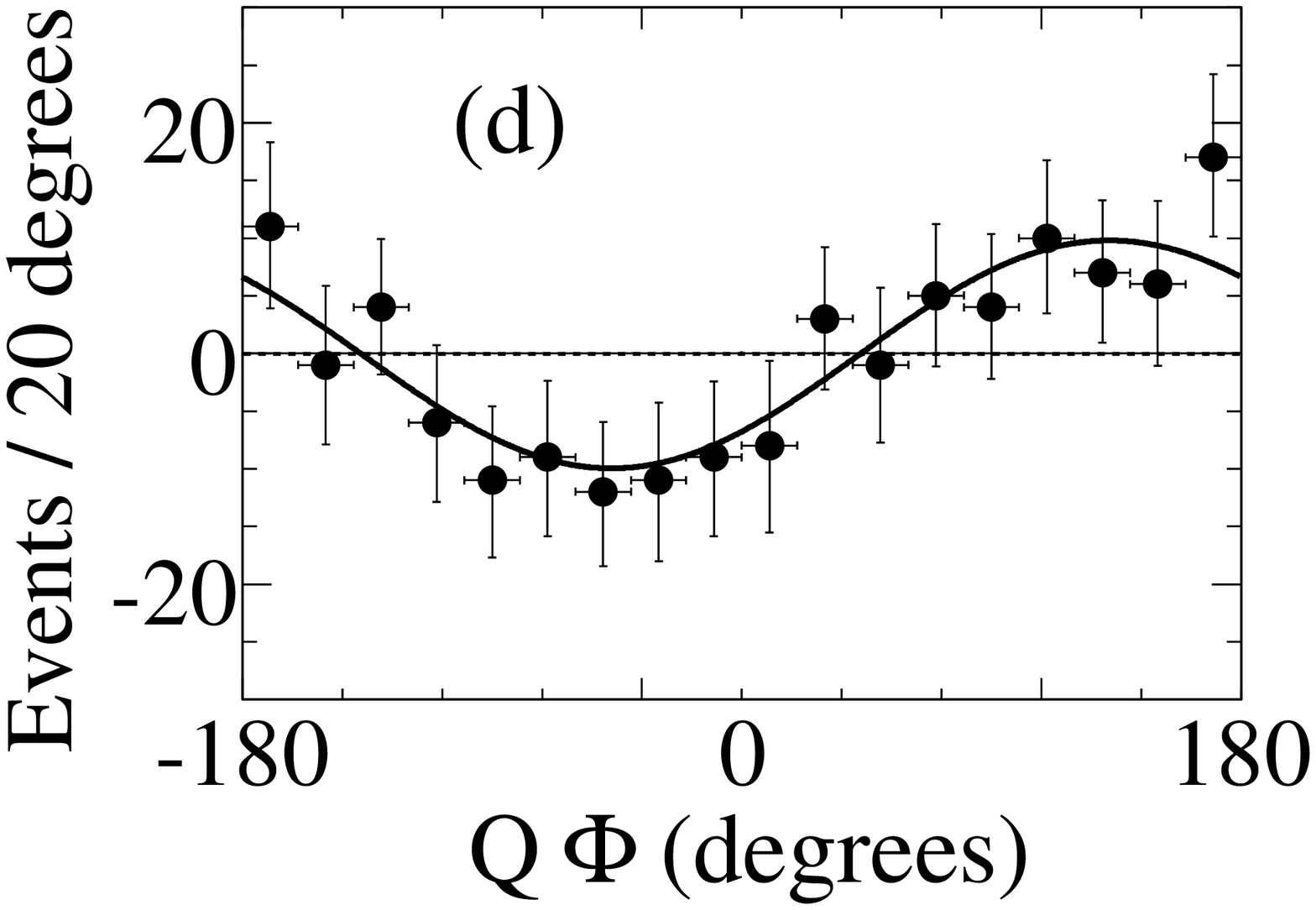}
}
\vspace{-0.3cm}
\caption{\label{fig:projection2} 
Projections onto the variables 
${\cal H}_1$ (a), ${\cal H}_2$ (b), $Q~\!\Phi$ (c), and
the differences between
the $Q~\!\Phi$ projections for events with 
${\cal H}_1~\!{\cal H}_2>0$ and with
${\cal H}_1~\!{\cal H}_2<0$ (d) 
for the signal $B^0\to\phi K^{*}(892)^0$ candidates
following the solid (dashed) line definitions in Fig.~\ref{fig:projection1}.
The $D^\pm_{(s)}$-meson veto causes the sharp acceptance dips seen in (a).
}
\end{figure}

\begin{figure}[b]
\centerline{
\setlength{\epsfxsize}{0.5\linewidth}\leavevmode\epsfbox{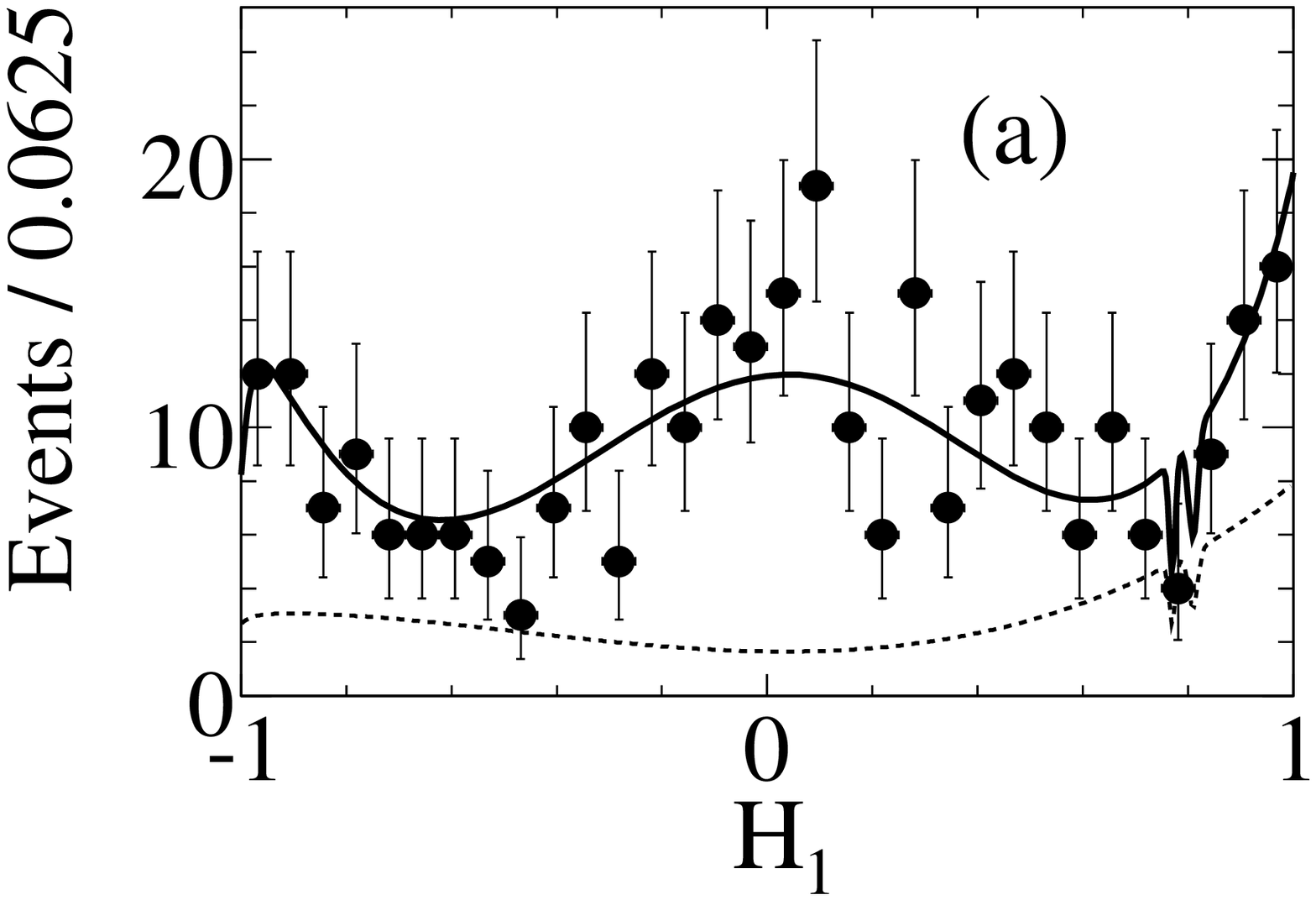}
\setlength{\epsfxsize}{0.5\linewidth}\leavevmode\epsfbox{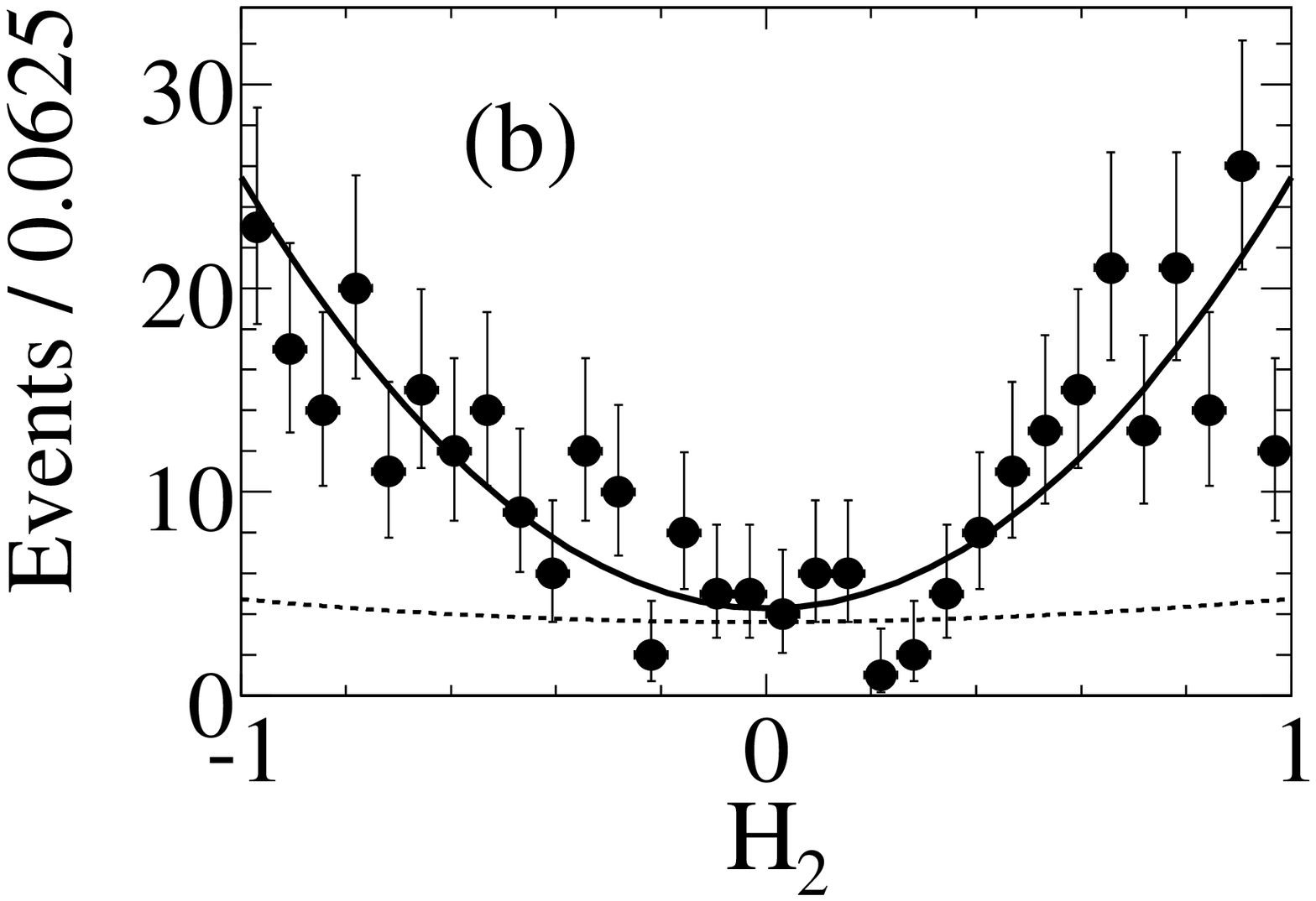}
}
\vspace{-0.3cm}
\caption{\label{fig:projection3} 
Same as Fig.~\ref{fig:projection2}(a) and \ref{fig:projection2}(b), 
but for the signal $B^0\to\phi K_2^{*}(1430)^0$ and
$\phi (K\pi)_0^{*0}$ candidates combined.
}
\end{figure}

The interference between the $J=1$ or $2$ and the
$S$-wave $(K\pi)$ contributions is modeled with 
the three terms $2{\cal R\rm e}(A_{J\lambda} A^*_{00})$ 
in Eq.~(\ref{eq:helicityfull}) with the 
four-dimensional angular and $m_{K\!\pi}$ dependence.
It has been shown in the decays $B^0\to J/\psi (K\pi)^{*0}_0$
and $B^+\to\pi^+(K\pi)^{*0}_0$~\cite{jpsikpi} that the amplitude 
behavior is consistent with that observed by LASS except for 
a constant phase shift. We allow an unconstrained overall shift,
again with the index $J$ suppressed,
$(\delta_0+\Delta\delta_0\times Q)$ between the LASS amplitude 
phase and either the vector ($J=1$) or the tensor ($J=2$) 
resonance amplitude phase.

The parameters {\boldmath$\xi$} describe the 
background or the remaining signal PDFs. 
They are left free to vary in the fit for the combinatorial 
background or are fixed to the values extracted from 
Monte Carlo (MC) simulation~\cite{geant} and calibration 
$B$-decay channels for the exclusive $B$ decays. 
We use a sum of Gaussian functions 
for the parameterization of the signal PDFs 
for $\Delta E$, $m_{\rm{ES}}$, and ${\cal F}$.
For the combinatorial background, we use polynomials,
except for $m_{\rm{ES}}$ and ${\cal F}$ distributions
which are parameterized by an empirical phase-space 
function and by Gaussian functions, respectively.
Resonance production occurs in the background and 
is taken into account in the PDF.

\begingroup
\begin{table*}[t]
\caption
{\label{tab:results1}
Fit results for each $m_{K\!\pi}$ range and signal component:
the reconstruction efficiency $\varepsilon_{\rm reco}$ obtained
from MC simulation; the total efficiency $\varepsilon$, 
including the daughter branching fractions~\cite{pdg2006}; 
the number of signal events $n_{\rm sig}$; statistical 
significance (${\cal S}$) of the signal; the branching fraction ${\cal B}$;
and the flavor asymmetry ${\cal A}_{\CP}$.
The branching fraction ${\cal B}(B^0\to\phi (K\pi)^{*0}_0$) refers to the coherent 
sum $|A_\text{res}+A_\text{non-res}|^2$
of resonant and nonresonant $J^P=0^+$ $K\pi$ components~\cite{Aston:1987ir} and 
is quoted for $m_{K\!\pi}<1.6$~GeV, while the ${\cal B}(B^0\to\phi K_0^{*}(1430)^0)$ 
is derived from it by integrating separately the Breit-Wigner formula of the 
resonant $|A_\text{res}|^2$
$K\pi$ component~\cite{Aston:1987ir} without $m_{K\!\pi}$ restriction.
The systematic errors are quoted last.
}
\begin{center}
\begin{ruledtabular}
\setlength{\extrarowheight}{1.5pt}
\begin{tabular}{ccccccccc}
\vspace{-3mm}&&&\\
Mode & $m_{K\!\pi}$ (GeV) & $m_{K\!\pi}$ model 
 & $\varepsilon_{\rm reco}$ (\%) & $\varepsilon$ (\%) & $n_{\rm sig}$ (events)
 & ${\cal S}$ ($\sigma$) & ${\cal B}$  ($10^{-6}$) & ${\cal A}_{\CP}$ \cr
\vspace{-3mm} & & & & \\
\hline
\vspace{-3mm} & & & &\\
 $\phi K^{*}(892)^0$ 
 & 0.75 -- 1.05 & Breit-Wigner~\cite{pdg2006} & $35.0\pm{1.7}$ & $11.5\pm 0.6$  & $406\pm 29\pm 15$
 & $21.0$ & $9.2\pm{0.7}\pm 0.6$ & $-0.03\pm{0.07}\pm 0.03$  
\\
\vspace{-3mm} & & & & \\
 $\phi K_2^{*}(1430)^0$ 
 & 1.13 -- 1.53 & Breit-Wigner~\cite{pdg2006} &  $27.1\pm{1.3}$ & $4.4\pm 0.2$  & $133\pm{19}\pm 7$
  & $9.7$ & $7.8\pm{1.1}\pm 0.6$ & $-0.12\pm{0.14}\pm 0.04$  
\\
\vspace{-3mm} & & & & \\
 $\phi (K\pi)^{*0}_0$ 
 & 1.13 -- 1.53 & LASS~\cite{Aston:1987ir} &  $23.4\pm{1.1}$ & $7.7\pm 0.3$  & $147\pm{23}\pm 7$
  & 9.8 & $5.0\pm{0.8}\pm 0.3$ & $+0.17\pm{0.15}\pm 0.03$  
\\
\vspace{-2mm} & & & & \\
 $\phi K_0^{*}(1430)^0$ 
 &  & Breit-Wigner~\cite{Aston:1987ir} &  &  &  & 
 & $4.6\pm{0.7}\pm 0.6$ & 
\\
\vspace{-3mm} & & & &\\
\end{tabular}
\end{ruledtabular}
\end{center}
\end{table*}
\endgroup


We observe a nonzero yield with more than 9$\sigma$ significance,
including systematic uncertainties, in each of the three 
$B^0\to\phi K^{*0}$ decay modes.
The significance is defined as the square root of the change in 
$2\ln{\cal L}$ when the yield is constrained to zero in the 
likelihood ${\cal L}$.
In Figs.~\ref{fig:projection1}--\ref{fig:projection3}
we show projections onto the variables.
In Tables~\ref{tab:results1} and~\ref{tab:results2}
the $n_{j}$, ${\cal A}_{j}$, and 
{\boldmath$\zeta$}~$\equiv\{f_L$, $f_{\perp}$, $\phi_{\parallel}$, 
$\phi_{\perp}$, $\delta_0$, ${\cal A}_{C\!P}^0$, ${\cal A}_{C\!P}^{\perp}$, 
$\Delta \phi_{\parallel}$, $\Delta \phi_{\perp}$, $\Delta\delta_0$\}
parameters of the $B^0\to\phi K^{*}(892)^0$ decay 
or the $\phi K_2^{*}(1430)^0$ and $\phi(K\pi)_0^{*0}$ decays
are obtained from the fit in the lower or higher
$m_{K\!\pi}$ range, respectively.


The nonresonant $K^+K^-$ contribution under the 
$\phi$ is accounted for with the $B^0\to f_0 K^{*0}$ category.
Its yield is consistent with zero in the higher $m_{K\!\pi}$ 
range and is $89\pm18$ events in the lower $m_{K\!\pi}$ range. 
The uncertainties due to $m_{K\!\Kbar}$ interference are 
estimated with the samples generated according to the observed 
$K^+K^-$ intensity and with various interference phases analogous 
to $\delta_0$ in $K\pi$. These are the dominant systematic 
errors for the {\boldmath$\zeta$} parameters of the 
$B^0\to\phi K^{*}(892)^0$ decay.

We vary those parameters in {\boldmath$\xi$} not used 
to model combinatoric background within their 
uncertainties and derive the associated systematic errors.
We allow for the flavor-dependent acceptance function
and the reconstruction efficiency in the study of asymmetries.
The biases from the finite resolution of the angle measurement, 
the dilution due to the presence of fake combinations, 
or other imperfections in the signal PDF model are estimated 
with MC simulation. Additional systematic uncertainty originates 
from $B$ background, where we estimate that only a few events 
can fake the signal.
The systematic errors in efficiencies are dominated 
by those in particle identification and track finding.
Other systematic effects arise from event-selection criteria, 
$\phi$ and $K^{*0}$ branching fractions, and number of $B$ mesons.


In the lower $m_{K\!\pi}$ range the yield of the 
$\phi({K\pi})^{*0}_0$ contribution is $60^{+17}_{-14}$ events 
with the statistical significance of 7.9$\sigma$,
including the interference term.
The dependence of the interference on the $K\pi$ invariant 
mass~\cite{Aston:1987ir, jpsikpi} allows us to reject the other 
solution near ($2\pi-\phi_{\parallel},\pi-\phi_{\perp}$)
relative to that in Table~\ref{tab:results2} 
for the $B^0\to\phi K^{*}(892)^0$ decay
with significance of 5.4$\sigma$, including systematic uncertainties.
We also resolve this ambiguity with statistical significance 
of more than 4$\sigma$ with the $\Bbar^0$ or $B^0$ decays independently.
Because of the low significance of our measured 
$f_\parallel=(1-f_L-f_\perp)$ (2.9$\sigma$)
and $f_\perp$ (1.6$\sigma$) 
in the $B^0\to\phi K^{*}_2(1430)^0$ decay
we have insufficient information to constrain $\phi_{\parallel}$ 
and $\phi_{\perp}$ at higher significance and to measure five 
asymmetries, and so we fix these asymmetry parameters to zero
in the fit in the higher $m_{K\!\pi}$ range.

The $(V-A)$ structure of the weak interactions and the
$s$-quark spin flip suppression in the diagram in Fig.~\ref{fig:decay}~(a)
suggest $|A_{0}|\gg|A_{+1}|\gg|A_{-1}|$~\cite{bvv1, newtheory}. 
This expectation is consistent with our measurements 
in the vector-tensor $B^0\to\phi K_2^{*}(1430)^0$ decay, 
but disagrees with our observed vector-vector polarization.
In the $B^0\to\phi K^{*}(892)^0$ decay we obtain the 
solution ${\phi_\parallel}\simeq{\phi_\perp}$ without
discrete ambiguities. Combined with the approximate solution
$f_L\simeq1/2$ and $f_\perp\simeq(1-f_L)/2$, this results
in the approximate decay amplitude hierarchy 
$|A_{0}|\simeq|A_{+1}|\gg|A_{-1}|$
(and $|\Abar_{0}|\simeq|\Abar_{-1}|\gg|\Abar_{+1}|$).

We find more than 5$\sigma$ (4$\sigma$) deviation, including
systematic uncertainties, of ${\phi_\perp} ({\phi_\parallel})$
from either $\pi$ or zero in the
$B^0\to\phi K^{*}(892)^0$ decay, indicating the presence of 
final-state interactions (FSI) not accounted for in naive factorization.
The effect of FSI is evident in the phase shift of the cosine
distribution in Fig.~\ref{fig:projection2} (d).

Our measurements of eight $C\!P$-violating 
parameters rule out a significant part of the
physical region and are consistent with no
$C\!P$-violation in this decay.
Significant nonzero $C\!P$-violating parameters
would indicate the presence of new amplitudes
with different weak phases.
The $\Delta{\phi_\perp}$ and $\Delta{\phi_\parallel}$ are particularly 
interesting due to sensitivity to the weak phases of the 
amplitudes without hadronic uncertainties~\cite{bvvreview2006},
such as the relative weak phases of $A_{+1}$ and $A_0$,
while the $C\!P$-violating $\Delta\delta_0$ parameter represents
potential differences of weak phases among decay modes.

In summary, we have performed an amplitude analysis and searched 
for $C\!P$-violation in the angular distribution with the 
$B^0\to\phi K^{*0}$ decays with the tensor, vector, and scalar $K^{*0}$.
Our results are summarized in Tables~\ref{tab:results1} 
and~\ref{tab:results2} and supersede corresponding measurements 
in Ref.~\cite{babar:vv}.
Our vector-tensor results are in agreement with quark 
spin flip suppression and $A_0$ amplitude dominance, 
whereas the vector-vector mode contains substantial 
$A_{+1}$ amplitude from a presently unknown source 
either within or beyond the standard model~\cite{newtheory}.

\begingroup
\begin{table}[t]
\caption{\label{tab:results2}
Summary of polarization results. 
The dominant fit correlation coefficients (${\cal C}$) 
are presented for the $\phi K^{*}(892)^0$ mode where 
we show correlations of ${\delta_0}$  with ${\phi_\parallel}/{\phi_\perp}$
and of ${\Delta\delta_0}$ with ${\Delta\phi_\parallel}/{\Delta\phi_\perp}$.
For the $\phi K_2^{*}(1430)^0$ mode, the dominant values of ${\cal C}$ are 
$32\%$ for $({\delta_0},{\phi_\parallel})$ 
and $26\%$ for $({\phi_\parallel}, {\phi_\perp})$.
}
\begin{center}
{
\begin{ruledtabular}
\setlength{\extrarowheight}{1.5pt}
\begin{tabular}{cccc}
\vspace{-3mm} & & \\
 & $B^0\to\phi K_2^{*}(1430)^0$ & $B^0\to\phi K^{*}(892)^0$  & ${\cal C}$ 
\cr
\vspace{-3mm} & & & \\
\hline
\vspace{-3mm} & & & \\
  ${f_L}$  
 & $0.853^{+0.061}_{-0.069}\pm 0.036$  
 & $0.506\pm{0.040}\pm 0.015$  
 & \multirow{2}{13mm}{~{\Large\}}$-53\%$}
\cr
\vspace{-3mm} & & & \\
  ${f_\perp}$ 
 & $0.045^{+0.049}_{-0.040}\pm 0.013$  
 & $0.227\pm{0.038}\pm 0.013$  
 & 
\cr
\vspace{-3mm} & & & \\
  ${\phi_\parallel}$ 
 & $2.90\pm 0.39\pm 0.06$ 
 & $2.31\pm 0.14\pm 0.08$ 
 &  \multirow{2}{13mm}{~{\Large\}}~$61\%$}
\cr
\vspace{-3mm} & & & \\
  ${\phi_\perp}$  
 & $5.72^{+0.55}_{-0.87}\pm 0.11$ 
 & $2.24\pm 0.15\pm 0.09$    
 &  
\cr
\vspace{-3mm} & & & \\
  ${\delta_0}$  
 & $3.54^{+0.12}_{-0.14}\pm 0.06$  
 & $2.78\pm 0.17\pm 0.09$    
 & $37\%/27\%$
\cr
\vspace{-3mm} & & & \\
  ${\cal A}_{C\!P}^0$ 
 & {\boldmath $\cdot\cdot\cdot$} 
 & $-0.03\pm{0.08}\pm 0.02$  
 &  \multirow{2}{13mm}{~{\Large\}}$-51\%$}
\cr
\vspace{-3mm} & & & \\
  ${\cal A}_{C\!P}^{\perp}$ 
 & {\boldmath $\cdot\cdot\cdot$}
 & $-0.03\pm 0.16\pm 0.05$   
 &  
\cr
\vspace{-3mm} & & & \\
  $\Delta \phi_{\parallel}$  
 & {\boldmath $\cdot\cdot\cdot$}
 & $+0.24\pm 0.14\pm 0.08$ 
 &  \multirow{2}{13mm}{~{\Large\}}~$61\%$}
\cr
\vspace{-3mm} & & & \\
  $\Delta \phi_{\perp}$  
 & {\boldmath $\cdot\cdot\cdot$}
 & $+0.19\pm 0.15\pm 0.08$   
 & 
\cr
\vspace{-3mm} & & & \\
  ${\Delta\delta_0}$  
 & {\boldmath $\cdot\cdot\cdot$}
 & $+0.21\pm 0.17\pm 0.08$    
 & $37\%/27\%$
\cr
\vspace{-3mm} & & & \\
\end{tabular}
\end{ruledtabular}
}
\end{center}
\end{table}
\endgroup


We are grateful for the excellent luminosity and machine conditions
provided by our \pep2\ colleagues,
and for the substantial dedicated effort from
the computing organizations that support \babar.
The collaborating institutions wish to thank
SLAC for its support and kind hospitality.
This work is supported by
DOE
and NSF (USA),
NSERC (Canada),
IHEP (China),
CEA and
CNRS-IN2P3
(France),
BMBF and DFG
(Germany),
INFN (Italy),
FOM (The Netherlands),
NFR (Norway),
MIST (Russia),
MEC (Spain), and
PPARC (United Kingdom).
Individuals have received support from the
Marie Curie EIF (European Union) and
the A.~P.~Sloan Foundation.


\bibliographystyle{h-physrev2-original}   

\begin{thebibliography}{99}

\bibitem{bvv1}
A.~Ali {\it et al.}, Z.\ Phys.\ C {\bf 1}, 269 (1979); 
G.~Valencia, Phys.\ Rev.\ D {\bf 39}, 3339 (1989);
G. Kramer and W.F. Palmer, Phys.\ Rev.\ D {\bf 45}, 193 (1992);
H.-Y.~Cheng and K.-C.~Yang, Phys.\ Lett.\ B {\bf 511}, 40 (2001);
C.-H. Chen {\it et al.}, Phys.\ Rev.\ D {\bf 66}, 054013 (2002);
M.~Suzuki, Phys.\ Rev.\ D {\bf 66}, 054018 (2002);
A.~Datta and D.~London, Int.\ J.\ Mod.\ Phys.\ A {\bf 19}, 2505 (2004).

\bibitem{bvvreview2006}
A.~V.~Gritsan and J.~G.~Smith, ``Polarization in $B$ Decays''
review in~\cite{pdg2006}, J. Phys. G33, 833 (2006).

\bibitem{pdg2006}
Particle Data Group,  W.-M. Yao {\it et al.}, J. Phys. G33, 1 (2006).

\bibitem{babar:vv}
$\babar$ Collaboration, B.~Aubert {\it et al.},
Phys.\ Rev.\ Lett.\ {\bf 91}, 171802 (2003); 
Phys.\ Rev.\ Lett.\  {\bf 93}, 231804 (2004).

\bibitem{belle:phikst}
Belle Collaboration, K.-F. Chen {\it et al.},
Phys.\ Rev.\ Lett.\ {\bf 91}, 201801 (2003); 
Phys.\ Rev.\ Lett.\ {\bf 94}, 221804 (2005).

\bibitem{newtheory}
A.~L.~Kagan, Phys.\ Lett.\ B {\bf 601}, 151 (2004);
Y.~Grossman,  Int.\ J.\ Mod.\ Phys.\ A {\bf 19}, 907 (2004);
C.~W.~Bauer {\it et al.}, Phys.\ Rev.\ D {\bf 70}, 054015 (2004);
P.~Colangelo {\it et al.}, Phys.\ Lett.\ B {\bf 597}, 291 (2004);
M.~Ladisa {\it et al.}, Phys.\ Rev.\ D {\bf 70}, 114025 (2004);
E.~Alvarez {\it et al.}, Phys.\ Rev.\ D {\bf 70}, 115014 (2004);
H.~Y.~Cheng {\it et al.}, Phys.\ Rev.\ D {\bf 71}, 014030 (2005);
H.~n.~Li and S.~Mishima, Phys.\ Rev.\ D {\bf 71}, 054025 (2005);
P.~K.~Das and K.~C.~Yang, Phys.\ Rev.\ D {\bf 71}, 094002 (2005);
C.~H.~Chen and C.~Q.~Geng, Phys.\ Rev.\ D {\bf 71}, 115004 (2005);
Y.~D.~Yang {\it et al.}, Phys.\ Rev.\ D {\bf 72}, 015009 (2005);
K.~C.~Yang, Phys.\ Rev.\ D {\bf 72}, 034009 (2005);
C.~S.~Huang {\it et al.}, Phys.\ Rev.\ D {\bf 73}, 034026 (2006);
M.~Beneke {\it et al.}, Phys.\ Rev.\ Lett.\  {\bf 96}, 141801 (2006);
C.~H.~Chen and H.~Hatanaka, Phys.\ Rev.\ D {\bf 73}, 075003 (2006).

\bibitem{Aston:1987ir}
LASS Collaboration,  D.~Aston {\it et al.},
Nucl.\ Phys.\ B {\bf 296}, 493 (1988);
W.~M.~Dunwoodie, private communications.

\bibitem{jpsikpi}
$\babar$ Collaboration, B.~Aubert {\it et al.},
Phys.\ Rev.\ D {\bf 71}, 032005 (2005);
Phys.\ Rev.\ D {\bf 72}, 072003 (2005).

\bibitem{babar}
\babar\ Collaboration, B.~Aubert {\it et al.},
{Nucl.\ Instr.\ Meth.\xspace} A {\bf 479}, 1 (2002).

\bibitem{bigPRD}
$\babar$ Collaboration, B.~Aubert {\it et al.},
Phys.\ Rev.\ D {\bf 70}, 032006 (2004).

\bibitem{f0mass}
E791 Collaboration, E. M. Aitala {\it et al.},
Phys. Rev. Lett. {\bf 86}, 765 (2001).

\bibitem{geant} S.~Agostinelli {\it et al.},
{Nucl.\ Instr.\ Meth.\xspace} A {\bf 506}, 250 (2003).

\end{thebibliography}

\end{document}